\documentstyle [12pt,a4]{article}
\newcommand{\be}{\begin{equation}}
\newcommand{\ee}{\end{equation}}
\newcommand{\bea}{\begin{eqnarray}}
\newcommand{\ena}{\end{eqnarray}}

\def\Li   {{\rm Li}}

\def\Re   {{\rm Re}}
\def\Im   {{\rm Im}}
\def\Ks   {\rlap/K}
\def\Ps   {\rlap/P}
\def\Qs   {\rlap/Q}

\def\(    {\left( }    \def\)   {\right) }
\def\[    {\left[}    \def\]   {\right] }
\def\txt #1 {\qquad {\rm #1} \qquad}
\def\lsim{\; \raise0.3ex\hbox{$<$\kern-0.75em\raise-1.1ex\hbox{$\sim$}}\; }
\def\gsim{\; \raise0.3ex\hbox{$>$\kern-0.75em\raise-1.1ex\hbox{$\sim$}}\; }
\def\@versim#1#2{\lower0.2ex\vbox{\baselineskip\z@skip\lineskip\z@skip
  \lineskiplimit\z@\ialign{$\m@th#1\hfil##\hfil$\crcr#2\crcr\sim\crcr}}}

\begin{document}
\begin{titlepage}
\begin{center}
\rightline{CERN-TH-7044/93}
\rightline{ENSLAPP-A-441/93}

\bigskip

{\LARGE \bf {Axion Emission}\\
\smallskip
{from Red Giants and White Dwarfs}}\\
\vspace{0.8cm}
{\large T. Altherr\footnote{On leave of absence from  L.A.P.P., BP110, F-74941
Annecy-le-Vieux Cedex, France}}\\
{\em Theory Division, CERN, CH-1211 Geneva 23, Switzerland}\\
\medskip
{\large E. Petitgirard}\\
{\em{Laboratoire de Physique Th\'eorique}}
{\small E}N{\large S}{\Large L}{\large A}P{\small P}
\footnote{URA 14-36 du CNRS, associ\'ee \`a l'E.N.S. de Lyon, et au L.A.P.P.
d'Annecy-le-Vieux.}\\
{\em B.P. 110, F-74941 Annecy-le-Vieux Cedex, France}\\
and\\
{\large T. del R\'\i o Gaztelurrutia}\\
{\em Departamento de F\'\i sica Aplicada}\\
{\em Escuela de Ingenieros, Alameda de Urquijo S/N}\\
{\em S-48013 Bilbao, Spain}\\[0.8cm]
\end{center}
\medskip

\centerline{ \bf{Abstract}}
Using thermal field theory methods, we recalculate axion emission from
dense plasmas. We study in particular the Primakoff and the
bremsstrahlung processes. The Primakoff rate is significantly suppressed at
high densities, when the electrons become relativistic. However, the bound
on the axion-photon coupling, $G<10^{-10}$ GeV, is unaffected, as it is
constrained by the evolution of HB stars, which have low densities. In
contradistinction, the same relativistic effects enhance the
bremsstrahlung processes. From the red giants and white dwarfs evolution,
we obtain a conservative bound on the axion-electron coupling,
$g_{ae} < 2\times 10^{-13}$.
\vfill

\leftline{CERN-TH-7044/93}
\leftline{October 93}
\end{titlepage}

\setcounter{footnote}{0}
\section{Introduction}

There are recent speculations that axions \cite{Kim} might play a role in
astrophysics \cite{IHG,ESH}. Indeed, the cooling rate of the white dwarf
G117-B15A has recently been measured and is faster than expected from any
conventional model. Axions with masses of the order of
$8(\pm 3)\times 10^{-3}/\cos^2\beta$ eV would provide an additional
cooling mechanism and reconcile models with the data \cite{IHG}.

On the other hand, it is well known that other constraints on axions
come from astrophysical systems such as red giants and SN1987A \cite{Raf0}.
A rather well-established bound comes from the evolution of red giants
\cite{Raf1}. The upper bound for DFSZ axions is
$m_a < 1.1\times 10^{-2}/\cos^2\beta$ eV, quite close to
the proposed solution for the cooling mechanism of G117-B15A.

The calculations of the axion emission rates in stellar cores
involve a quite complicated physics (screening and correlation effects).
They are based upon kinetic theory \cite{SI}. As a consequence of this
complexity, there has been a considerable amount of confusion in
identifying the correct physics for the photon propagation in dense media
\cite{FWY,CNP}. The best approach is due to Raffelt \cite{Raf1,Raf2}, and we
shall reproduce some of his results here. Indeed,
given the outstanding implications of the possible existence of axions, we
propose to recalculate the dominant processes of axion emission in red
giants and white dwarfs. We shall use a different
method, derived from thermal field theory \cite{Kap,LvW}. The advantage of this
method is that it already includes the classical limit, but allows also
systematic diagrammatic studies of the higher-order effects \cite{AK}.

The paper is organized as follows: in Sect.~2 we recall some of the basic
ingredients concerning the spectral density of photons at finite
temperature and density. In Sect.~3, we investigate the Primakoff process
in details. We confirm the previous bound on the axion-photon coupling
strength.
In Sect.~4, we calculate the axion emission from electron-nucleus
bremsstrahlung
through the exchange of transverse and longitudinal photons.
We discuss our results in Sect.~5.
{}From the red giants and white dwarfs evolution,
we obtain a conservative bound on the axion-electron coupling,
$g_{ae} < 2\times 10^{-13}$.

\section{The method}

The general framework and methods of gauge field theory at finite
temperature and density have been discussed extensively elsewhere
\cite{AK},
and we shall not repeat them here. Let us just recall that in the plasma core
of dense stars, matter effects are properly taken into account by the
finite temperature and density corrections to the polarization tensor
$\Pi(Q)$. In a heat bath, there exist two modes of propagation of
electromagnetic waves, transverse (T) and longitudinal (L).
In calculating an electromagnetic scattering, it is particularly
convenient to make use of the spectral density \cite{Pis}
\bea
  \rho_{\rm T,L}(Q) &=&{1\over \pi} \Re {i\over Q^2-\Pi_{\rm T,L}(Q)+i\epsilon}
\nonumber\\
                    &=& \delta(Q^2-\Re\Pi_{\rm T,L}(Q)) - \beta_{\rm T,L}(Q)
.\ena
Our convention is to use capital letters only for quadri-momenta, {\it
e.g.} $Q=(\omega,{\bf q})$.
Depending on whether the photon is time-like or space-like, the above
quantity describes two different modes:

i) If the photon is time-like ($Q^2>0$), the spectral density turns to a
$\delta$-function and describes a quasi-particle mode (plasmon).
This quasi-particle obeys a dispersion relation given,
in the non-relativistic limit, by $\omega^2=\omega_0^2+q^2$ for the
transverse mode, and by $\omega^2=\omega_0^2$ for the longitudinal one.
Similar, although much more involved, analytic expressions can be derived
in the ultra-relativistic limit. Recent works have shown that it is
possible to avoid the non/ultra-relativistic limit and get closed formulae
\cite{APR,BS}.
The plasmon frequency is given by $\omega_0^2=e^2N_e/m_e$ for
non-degenerate and non-relativistic plasmas and by $\omega_0^2=e^2N_e/\mu$
for degenerate plasmas ($\mu$ is the relativistic electron chemical potential,
$\mu=\sqrt{p_F^2+m_e^2}$). In the non-relativistic case, the longitudinal
dispersion relation crosses the light-cone at $q=\omega_0$, and eventually
continues for large values of $q$ until an imaginary part develops in
$\Pi_{\rm L}(Q)$. This peculiar feature of a space-like quasi-particle occurs
only in a medium. It will be particularly relevant for the Primakoff
effect, as we shall see in the next section.

ii) At leading order, the polarization tensor develops an imaginary part only
below the light-cone, for instance when $\omega<p_Fq/\mu$ in the degenerate
case. In the hard-loop approximation, the imaginary parts of the transverse
and longitudinal components are given
by
\bea
\displaystyle
\Im\Pi_{\rm T}(Q) &=& \alpha {1\over q} \int_m^\infty dE
\( \frac {\omega^2}{q^2} E^2 - p^2\)
    \theta(pq-\omega E) \[ n_F\( E-{\omega\over 2}\) -n_F\( E+{\omega\over
2}\) \]
\nonumber\\
\Im\Pi_{\rm L}(Q) &=& 2\alpha \( 1-\frac {\omega^2}{q^2}\) \frac {1}{q}
                \int_m^\infty dE\  E^2 \theta(pq-\omega E)
                \[ n_F\( E-{\omega\over 2}\) -n_F\( E+{\omega\over 2} \) \]
,\ena
where the contribution from antiparticles has been dropped.
The spectral density then turns to some kind of Lorentzian with a
complicated structure. One may notice that longitudinal and transverse
expressions are very similar in the
ultra-relativistic limit \cite{AK}, but not in the non-relativistic case
\cite{APR}. In the non-relativistic and non-degenerate case, we recover
the standard relation \cite{SI}
\be
\Im \Pi_{\rm L}(Q) = \sqrt{\pi} k_D^2 z \ e^{-z^2} \txt{with} z={\omega\over q}
\sqrt{m\over 2T}
,\ee
and $k_D$ is the Debye mass corresponding to the screening of static
electric fields. In non-degenerate plasmas, $k_D^2=e^2 N_e/T$,
and in degenerate plasmas, $k_D^2=(e^2/\pi^2)\mu p_F$.

As we shall see in the following, it is important to notice that screening
effects are absent for transverse modes, while there appears a mass term
into the denominator of the longitudinal spectral density. More precisely,
one has in the static limit
\bea
          \Re\Pi_{\rm T}(\omega\to 0,q) &=& k_D^2 {\omega^2\over q^2}
\nonumber\\
\txt{and} \Re\Pi_{\rm L}(\omega\to 0,q) &=& k_D^2.
\ena
Notice that $k_D$ also appears in the expression for the transverse
propagator. Because of the $\omega$ dependence, one may call this effect
a dynamical screening \cite{Pis}.

All the above equations have been calculated for the case of electrons or
protons. Similar expressions can be derived in case of ions, using scalar
electrodynamics instead of quantum electrodynamics. This is legitimate
as, in most scatterings, the photon does not see the inner structure
of the nucleus. Moreover, one does not need form factors as we only
consider the thermal corrections, which are ultraviolet-finite.
Equations~(3) and (4) are identical, except for the Debye mass, which becomes
$k_D^2 = Z^2 e^2 N_{ions}/T$.

\bigskip
The scattering rates are computed either using the cutting rules of Kobes
and Semenoff \cite{KS} in the real-time formalism \cite{LvW}, or by
analytic continuation in the imaginary-time formalism \cite{Wel}.
Many examples can be found in the literature, including those
discussed here, but mostly in the ultra-relativistic regime
\cite{AK,BS,Alt,BY}.

\section{The Primakoff effect}
In a previous work \cite{AK}, the axion emission
through photon exchange was already studied in detail for the case of an
ultra-relativistic plasma. This analysis can  be extended to the general
case of relativistic as well as non-degenerate plasmas.
As shown in \cite{AK}, a systematic study of
higher-order effects involves a resummation of a class of diagrams. This
resummation is best implemented within Thermal Field Theory. In this
approach, the axion emission rate is calculated through the discontinuity
of its self-energy. This procedure avoids the complication of using Feynman
scattering amplitudes, and yields the contribution
of several different processes to the axion production. Nevertheless, it
remains difficult to compare the results thus obtained  with kinetic theory
calculations \cite{Raf1}.

When we calculate the discontinuity of the axion self energy,
three different kinds of contributions naturally appear. Out of them, the
``pole-cut'' term is what we can best compare with the classical Primakoff
process. It involves the scattering of a time-like plasmon with
a target (which can either be an ion or an electron) through the exchange
of a space-like plasmon. To be able to compare our results with previous
calculations, we shall concentrate on the case where
the time-like photon is transverse, while the space-like one is
longitudinal. The equivalence between Primakoff and
pole-cut is exact when the photon energies are hard,
i.e. larger than the Debye mass or the plasmon frequency; it ceases to
be valid when the photons are soft, as the space-like photon is
affected by Landau damping.

In our calculation of the axion emission of stars, we shall first compute the
dominant pole-cut contribution, then discuss when the pole-pole
contribution becomes relevant, and finally find a complete expression
valid in the non-degenerate, non-relativistic case.

Following previous work \cite{AK}, we start from the expression for the
pole-cut contribution to the emission rate of axions
\be \Gamma_{PC}(K)=
{G^2\over 8\pi k} \int_{\omega_0}^\infty d\omega \int_{-1}^{+1} \sin\theta
d\theta\ q n_B(\omega)(1+n_B(\omega')) Z_2(Q) \beta(Q') g(Q,Q') ,\ee
where the same notation as in \cite{AK} is used ($Q=Q'+K$).
In the last expression, $Q'$ is the momentum of the space-like, longitudinal
plasmon, while $Q$ is the momentum of the transverse, on-shell photon (see
Fig.~1).
For a pseudoscalar coupling one has $g(Q,Q') =q^2 Q'^2\sin^2\phi$,
where $\phi$ is the
angle between $Q$ and $Q'$ \cite{Alt}.

The non-relativistic dispersion
relations for a transverse photon give
\be \omega^2 = q^2 + \omega_0^2  \txt{and}  Z_2(Q)=1 .\ee
In fact, the dispersion relation for the transverse photon in the
non-relativistic limit differs only very slightly from the ultrarelativistic
one \cite{APR,BS}.
We assume that, in the calculation of the pole-cut term, the approximation
of the dispersion relation
given above can be used in all regimes to a good accuracy.
We need also the `cut' contribution of the spectral density of the
longitudinal plasmon. As a first approximation we shall take
\be
\beta_{\rm L}(Q') \simeq -{1\over \pi} {\Im\Pi_{\rm L}(Q')
\over (q'^2+ k_D^2)^2}
,\ee
where the longitudinal plasmon is static.  As
the imaginary part of $\Pi_{\rm L}$ only exists for $\omega'< v_* q'$,
we expect this
approximation to hold well in the case of a non-relativistic plasma, where
the ion velocity $v_*$ is indeed very small, both in a degenerate and
in a non-degenerate electron gas.
The Debye mass appearing in the photon propagator is in fact the
sum of the different contributions to the screening (electrons, protons and
dominantly ions contributions).

The static limit allows a direct evaluation of
\be
\int d\omega' (1+n_B(\omega')) \Im\Pi_{\rm L} \simeq T \int {d\omega'\over
\omega'} \Im\Pi_{\rm L}\simeq \pi T k_D^2
.\ee
After performing the remaining integration on $\theta$, one
gets the final result for the pole-cut contribution to the emission rate
of axions
\be
\Gamma(K) = {G^2\over 8 \pi} T k_D^2 n_B(k) \sqrt{1-{\omega_0^2\over k^2}}
\[ {2k^2-\omega_0^2+k_D^2 \over 4k\sqrt{k^2-\omega_0^2}}
\ln{ { (\sqrt{k^2-\omega_0^2}+k)^2 + k_D^2
 \over (\sqrt{k^2-\omega_0^2}-k)^2 + k_D^2 } }  - 1 \]
.\ee
When the axion momentum $k$ is much larger than the plasmon frequency
$\omega_0$, the expression in square brackets simplifies to
\be
[\ldots] \to \[ {1\over 2} \( 1+{k_D^2 \over 2k^2} \)
                    \ln{\( 1 + {4k^2\over k_D^2}\) } - 1 \]
.\ee
The main approximation in the calculation of the pole-cut contribution is the
static limit for the longitudinal space-like plasmon. This approximation
should work well in a large variety of regimes, as it is the interaction
with the ions that contributes most to both the Debye length and the
imaginary part of $\Pi_{\rm L}$. The ions being always non-relativistic, our
approximation is well justified.

\bigskip
The pole-pole contribution corresponds to the decay of a
transverse photon into an axion and a plasmon.
When calculating it,
we must use the longitudinal plasmon dispersion relation. In the low-momentum
limit, this dispersion relation gives a constant energy
$\omega_0$, while in the large momentum limit, it tends asymptotically
to $\omega=v_* q$. Moreover, the Jacobian includes a factor $\omega-v_*q$,
which
suppresses
the process once the asymptote is reached. In the case of a very dense
plasma, the asymptote coincides with the light cone, and only $q$'s of the
order of or smaller than the frequency of the plasma
contribute. Nevertheless, the phase space increases considerably in the
case of non-relativistic, dilute plasmas, where $q$ can be as large as
$\omega/v_*$. Notice that in this case we have a space-like solution for
the longitudinal particle-like propagation in the plasma.

It is hard to find a good analytical approximation for the pole-pole
contribution in regions other than non-relativistic and non-degenerate.
Therefore, we perform the calculation numerically. We start again
from the results of \cite{AK},
\be \Gamma_{PP}(K) =
{G^2\over 8\pi k} \int_{\omega_0}^\infty d\omega n_B(\omega)(1+n_B(\omega'))
Z_2(Q')Z_2(Q) g(Q,Q') .\ee
Here, $\omega(q)$ and $Z_2(Q)$ are the same as in Eq.~(6), while the
longitudinal dispersion relations give \cite{APR}
\be q'^2 = {3\omega_0^2\over v_F^2}\[ 1-{1\over 2}{\omega'\over
v_F q'}\ln {\omega' + v_F q'\over \omega' - v_F q'}\]
\txt{and} Z_{2}(Q')={2\over 3}{q'^2\over
Q'^2}\( \omega'^2-v_F^2 q'^2\over \omega_0^2-\omega'^2+v_F^2 q'^2\) .\ee
We show the result of our numerical calculation in Fig.~2.
The pole-pole contribution gets large at low densities, and is indeed
dominated by the exchange of space-like longitudinal plasmons. The curve
follows remarkably the pole-cut term, that is Eq.~(9). At high densities,
though, the pole-pole is dominated by the exchange of time-like
longitudinal plasmons, and is still $\sim$ 50\% of the pole-cut rate.
However, in this region, the rate is dominated by the pole-cut term with
exchange of a space-like transverse plasmons (we shall comment more in
detail on that point below).

We have also plotted the production rate predicted by Raffelt. We see that
our prediction only agrees in the region of low densities. We shall argue
that this is consistent with the approximations used in both calculations.

In his calculation, Raffelt used a form factor different from the standard
Debye-H\"uckel one. His justification of this point relies in the
stationarity of the targets. The relevant time scale for a Primakoff-type event
to occur is the time it takes a photon to cross a region of the size
of the Debye length. If the incident photon is very energetic this time
will be $k_D^{-1}$, much smaller than $v_ek_D^{-1}$, the time an electron
takes to cross the same region. But whenever the temperature is of the
order of $\omega_0$, the velocity of the average plasmon excitation will be
much
smaller
than the speed of light, and the argument above will cease to be correct.
Indeed, we see from our calculation that when the densities grow, the
pole-cut term becomes dominant. In fact, it is easily proved  that our pole-cut
contribution is equivalent to the result that would be obtained using the
standard Debye-H\"uckel form factor in the calculations of \cite{Raf1}.

\bigskip
Within the non-relativistic, non-degenerate approximation, a
compact result for the axion emission rate can be deduced, which reproduces
exactly the result of ref. \cite{Raf1}.
If we consider both pole-pole and pole-cut contributions together, the
axion emission rate becomes
\be \Gamma(K) =
{G^2\over 8\pi k} \int_{\omega_0}^\infty d\omega \int_{-1}^{+1} \sin\theta
d\theta\ q\ n_B(\omega)(1+n_B(\omega')) Z_2(Q) \rho_{\rm L}(Q') g(Q,Q') ,\ee
where $\rho_{\rm L}$ is now the complete spectral density for the longitudinal
plasmon.
In terms of the electric permittivity $\epsilon_L$, one has \cite{SI}
\be \beta_{\rm L}(Q')=-{1\over Q'^2} \Im {1\over\epsilon_{\rm L}} \ee
In the classical limit, Eq.~(13) can be approximated to
\be \Gamma(K) =
{G^2\over 8\pi } k^2n_B(k)\int_{-\infty}^\infty d\omega' {T\over \omega'}
 \int_{-1}^{+1}
d(\cos\phi) \sin^2\phi\( -\Im {1\over\epsilon_{\rm L}}\)  .\ee
At this stage, it is quite convenient to use the Kramers-Kr\"onig sum rule
\cite{SI}
      \be \int_{-\infty}^{+\infty}~ {d\omega'\over\omega'}\Im
{1\over\epsilon_{\rm L}}=-\pi {k_D^2\over q'^2+k_D^2}
.\ee
We are left with the angular integration, which can be solved exactly,
leading to the final result:
\be
\Gamma(K) = {G^2\over 16 \pi} T k_D^2 n_B(k)
\[ \( 1+{k_D^2\over 4k^2}\) \ln\( 1+{4k^2\over k_D^2}\)  -1 \] ,\ee
which indeed coincides with ref. \cite{Raf1}. We notice that in the large-$k$
region, the sum rules are fulfilled by the pole-cut term, while in the low-$k$
region the pole-pole dominates.

\bigskip Let us now comment on two other points:

\noindent i)
The main approximation, which leads to our final result,
is the static limit in the
space-like photon propagator. It is obviously an overestimate. However,
it is certainly a very good approximation in the non-relativistic limit as
$\omega< q p/E = q \sqrt{3T/M} \simeq 0.003 q$ (for $^4$He). As axions
mostly originates from ion scatterings, we see that this approximation
is well justified.
\medskip

\noindent ii)
In Fig.~2, we also show the contribution coming from the exchange of
{\it transverse} photons. Unfortunately, we have not been  able to compute
analytically the axion emission rate in this case, except in a few limiting
cases. However, we can use the numerical codes developed previously
for the ultra-relativistic regime \cite{Alt,AK}. With the help of the results
of \cite{APR,BS}, the implementation of the general case of a degenerate gas
of electrons is straightforward. Notice that, in this case, the dominant
process comes from scatterings where the targets are not ions but electrons.
Still, the Debye mass that one has to use in the photon propagator (see
Eq.~(4)) is the sum of all the contributions.

The justification for neglecting transverse space-like
excitations is because of their usual small fluctuations in the
non-relativistic limit \cite{Raf2,APR}. However, and contrary to the
longitudinal case, these excitations are not screened and can therefore
contribute significantly when the Debye mass gets large. Also, as the
degeneracy of the electron gas increases, the spectral density of transverse
excitations becomes as large as the longitudinal ones. And this is
definitely the case in red giants, when the density reaches $\rho = 10^6$
g/cm$^3$ \cite{APR}.

The numerical results shown in Fig.~2 clearly demonstrate this feature.

\section{Bremsstrahlung}

Another important  contribution to the emission rate of
axions can be the bremsstrah\-lung process. In this case, a plasmon is
absorbed by an electron or ion, and an axion is emitted. In
contradistinction to a Primakoff-type process, bremsstrahlung occurs
via the fermion-axion coupling, $-ig\bar \Psi \gamma_5 \Psi \Phi_a$. A
derivative axion coupling was
introduced by Raffelt, to set right the
potentially incorrect consequences of the usual pseudoscalar coupling.
Nevertheless, in the case of the bremsstrahlung
process, it can easily be shown that both couplings can be used, and lead to
identical results.

In the case of ultra-degenerate plasmas of electrons, several
approximations can be performed. On the one hand, Pauli-blocking effects
imply that only electrons near the Fermi sphere contribute \cite{AK}.
 These electrons are very energetic, with energies of the order of
$\mu$, and the hard-loop approximation will be justified.
Another consequence of Pauli-blocking is the strong suppression of
the $e^-  e^-$ bremsstrahlung processes in comparison to $e^-$-ion
ones.

We shall be concerned with the imaginary part of the polarization tensor,
taking only into account the exchange with the nuclei in the photon
propagator. We only consider the space-like photon, as the time-like
photon is related with the Compton process, which has been shown to be
subleading \cite{Raf0}. In a strongly degenerate plasma, this is easy to
understand, as the plasmon density is automatically damped by a factor $e^{-
\omega_0/T}$.
Using the cutting rules of Kobes and Semenoff, the emission rate is
written as (see Fig.~3)
\bea
\Gamma(K)&=&{g_{ae}^2\over k} \int {d^4P\over (2\pi )^3}{d^4Q\over (2\pi)^3}
\( \theta(p_0-k_0)-n_F(p_0-k_0)\) \delta \( (P-K)^2-m_e^2\)
\nonumber\\ &&\times \( \theta(q_0-p_0)-n_F(p_0-q_0)\) \delta \(
(P-Q)^2-m_e^2\)  \nonumber\\ &&\times \( \theta(-q_0)+n_B(q_0)\) \beta (Q) ~
\sum_{spins} |{\cal A}|^2,
\ena
where the amplitude is given by
\be
{\cal A} =  \quad\epsilon_{\mu}^{s}(Q)~\bar u^{r'}(P-K)\[ \gamma^5
{\Ps +m_e \over P^2-m_e^2}\gamma^{\mu} +\gamma^{\mu} {\Ps -\Qs -\Ks +m_e \over
(P-Q-K)^2-m_e^2}\gamma^5\] u^{r}(P-Q).
\ee
After summing over all spins and photon polarizations, we can split the total
squared matrix element into its longitudinal and transverse parts.
Using the hard-loop approximation, we obtain,
\bea
\sum_{spins}   | {\cal A}~ |^2_{\rm T} &=& {-2\over (P\cdot  K)^2}\[
    2(K\cdot  Q)^2+Q^2 k^2(-1+({\bf \hat k \cdot\hat q})^2)\] ; \\ \nonumber
\sum_{spins}  | {\cal A}~ |^2_{\rm L} &=& {-2k^2Q^2\over (P\cdot K)^2}\[
1-({\bf
\hat k \cdot\hat q})^2\] .
\ena
In a very dense plasma, Fermi distribution functions can be well
approximated by step functions. This approximation yields the following
constraints on the energies
\be \mu +k_0 < p_0 < \mu + q_0, \ee
and allows us to write
\be n_B (q_0) n_F(p_0-q_0) (1-n_F(p_0-k_0))
\simeq
n_B(q_0)(1+n_B(q_0-k_0))(q_0-k_0)\delta(p_0-\mu). \ee
At this stage, the
integrations over $p_0$ and $p$ are easily solved, leading to
\bea  \Gamma_{T,L}(K)&=&{g_{ae}^2\over k} p_F
\int {d^4Q\over (2\pi )^4} \int{d
\Omega_p\over 8\pi^2}~ \beta_{T,L} (Q)
n_B(q_0)(1+n_B(q_0-k_0))\nonumber \\ &&\times(q_0-k_0)
\delta \( (P-Q)^2-m_e^2\) \sum_{spins} | {\cal A}~ |^2_{T,L}. \ena

We shall first calculate the transverse contribution to the emission
rate, and proceed later to the longitudinal case. Before performing the
angular integrations, let us have a look at the spectral densities of the
exchanged photons. The exchange with the nuclei involves the space-like
contribution of the spectral density (Eq.~(1)). In the case of heavy,
non-relativistic ions we can use the Boltzman statistical factor in the
calculation of $\Im \Pi_{\rm T}$, and we obtain:
\be \Im \Pi_{\rm T} = -\sqrt{\pi \over 6} k_D^2 v_* {T\over q}
e^{{ {-q_0\over2T}}}n_B^{-1}(q_0) e^{{  -{q_0^2\over
q^2}{3\over 2 v^2_*}}}
,\ee
where $v^2_*=3T/M \ll 1$ is the average
speed of ions and $M$ is the mass of the nuclei. The exponential factor
allows us to approximate $q_0\ll q$. Moreover, if we recall that
screening effects are absent for transverse modes, we see that $\Re
\Pi_{\rm T}
\ll q^2$ and $\Im \Pi _{\rm T} \ll q^2$.
Therefore the transverse spectral density has the following form
\be
   \beta_{\rm T} (Q) \simeq - {1\over \pi}{ \Im \Pi_{\rm T}(Q)\over q^4}
,\ee
which just corresponds to using bare propagators.
Performing the angular integrations is not a trivial task, since three
combined angles appear, namely ${\bf \hat k \cdot\hat q}$,
${\bf \hat k \cdot\hat p}$ and ${\bf \hat p \cdot\hat q}$. To ease the
calculation, a usual trick, based on the isotropy of the emission rate,
can be used,
\be \Gamma (K) = {1 \over 4\pi} \int d \Omega_{k'}\Gamma(K') .\ee
Notice that if ${\bf \hat k \cdot\hat p}$ for fixed ${\bf p}$ is not
constrained, ${\bf \hat q \cdot\hat p}$ for fixed ${\bf q}$ is on the
other hand bounded.

We arrive at an expression for $\Gamma$ where only one
integral is left unsolved,
\bea \Gamma_{\rm T}(K)&=&{g_{ae}^2\over 192 \pi^4} {k_D^2 v_*^2 T\over k\ p_F^2
}
 \[ {\mu\over p_F}\ln{\mu+p_F\over\mu-p_F}+2\( {p_F^2\over
m_e^2}-1 \) \] \nonumber\\
&&\times \int_0^{\infty} d q_0 ~(1+n_B(q_0)){q_0\over
q_0+k}~e^{-{ {q_0\over2T}}}~e^{-{{k\over2T}}}
.\ena
Finally, we can consider the energy loss rate per volume, related to the
emission rate by
\be
Q_{\rm T}={1\over 2\pi^2}\int_0^{\infty}dk k^3\Gamma_{\rm T}(K)
.\ee
{}From the last expression of the emission rate, we can deduce the value of the
energy loss rate for a transverse photon exchanged,
\bea
Q_{\rm T}&=&{g_{ae}^2\over 48\sqrt 2 \pi^6}{k_D^2 v_*^2 T^5\over p_F^2}
 \[ {\mu\over
p_F}\ln{\mu+p_F\over\mu-p_F}+2\( {p_F^2\over m_e^2}-1 \) \]
\nonumber \\
&&\times\( {\pi^2 \over 4}-{5\over 3} \zeta(3) + {1\over 8} \int
_0^{\infty} dx~x^3 (1+n_B(x)) ~\Li(x/2)\) .\ena

\bigskip

We now turn to the longitudinal case. The imaginary part of the
polarization tensor for heavy, non-relativistic ions is,
\be \Im \Pi_{\rm L} (Q) = \sqrt{3 \pi \over 2}{k_D^2\over v_*}{T\over q}
{}~n_B^{-1}(q_0) e^{{ -{q_0^2\over q^2}{3\over 2 v^{*2}}}}
.\ee
Its structure is such that it allows us to approximate
$q_0\ll q$, as in the transverse case.
Then, $\Re\Pi_{\rm L}\simeq k_D^2$, where $k_D$ is the Debye
mass. Moreover, $\Im \Pi_{\rm L}$ is negligible compared with $(Q^2-\Re
\Pi_{\rm L}(Q))^2$ in the denominator of the spectral density, and the
spectral density can be written as
\be
  \beta _{\rm L}(Q)\simeq -{1\over\pi}{\Im\Pi_{\rm L}(Q)\over\( q^2+k^2_D\) ^2}
.\ee
The calculation of the emission rate and energy loss rate, although
more involved, proceeds in the same manner as in the transverse case.
The final result is
\bea
  \Gamma_{\rm L}(K)&=&{g_{ae}^2\over 32\pi^5} \sqrt{3\pi \over 2}{k_D^2 T\over
k p_F^2
  v_*}
  \[ 2\( {p_F^2\over m_e^2}+1 \) -{\mu\over
  p_F}\ln{\mu+p_F\over\mu-p_F} \] \nonumber\\
&&\times\int_k^{\infty} d q_0~
(1+n_B(q_0-k_0))~(q_0-k_0)e^{{ -{q_0\over2T}}}
 \int_{q_0}^{\infty}{q^2 dq\over \( q^2+k^2_D\) ^2}
{}~e^{{ -{q_0^2\over q^2}{3\over 2v_*^2}}}
.\ena
The calculation of the energy loss rate may be easier, using the Parseval
theorem between functions and their Fourier transforms:
\be
\int_{q_0}^{\infty} dq {q^2\over (q^2+k_D^2)^2}e^{{-{q_0^2\over q^2}
{3\over 2v_*^2}}} \simeq {\sqrt{\pi}\over 4k_D}\int_0^{\infty}dx
e^{-\sqrt{3\over 2}{x q_0\over k_D v_*}}
\( 1+\sqrt{3\over 2}{x q_0\over k_D v_*}\)
.\ee
Performing first the integration over $q_0$, and then that over $k$, an exact
expression is obtained for the energy loss rate, with the remaining simple
integration:
\bea
Q_{\rm L}&=&{g_{ae}^2\over 128\pi^6} {k_D^2 T^5\over p_F^2}
\[ 2\( {p_F^2\over m_e^2}+1 \) -{\mu\over
  p_F}\ln{\mu+p_F\over\mu-p_F} \]  \nonumber\\
&&\times\int_0^{\infty} dx e^{-{x^2\over 4}}r_* F^{-3}(x) \[ \( \zeta(2,F(x)+1)
+F^{-2}(x) \) \( \sqrt{3\over 2}+{9\over 2}xr_*F^{-1}(x) \) \right. \nonumber\\
&& \left. +\ 3 xr_*\( \zeta(3,F(x)+1)+F^{-3}(x) \) \]
,\ena
with $F(x)={1\over 2}(1+\sqrt6 r_*x)$, $r_*={T\over k_Dv_*}$, and the
generalized $\zeta$-function $\zeta(a,b)$.

The longitudinal energy loss rate is obviously much larger than the transverse
one, by a factor $1/v_*^2$. The result can be decomposed into an
electronic part and a remaining integral, which contains the physics of
Landau damping with the nuclei.
This integral depends on a single parameter, $k_D v_*/T$, which
can vary between 1 and 20 for the physical cases of interest. Numerically,
we find that this integral is about 11., and is almost independent of
$r_*$, within 6\% accuracy, so that we might as well take its value for
$k_D=0$. This
seems to be in agreement with previous works, where it was found that
the Debye screening is negligible \cite{Raf1}. The temperature dependence of
the  rate is therefore as $T^4$.

\section{Discussion}

Using thermal field theory methods, we have calculated the axion emission
rate through the Primakoff mechanism. We have found that relativistic
effects strongly suppress the rate at high densities, compared with the
non-relativistic approximation. Nevertheless, the axion-photon coupling
constraint, $G < 10^{-10}$ GeV$^{-1}$ \cite{Raf1,Raf2}, is not affected by
these results. Indeed, this constraint comes from the evolution of HB
stars, where the density is about $\rho\simeq 10^4$ g/cm$^3$, and where the
electrons are clearly non-relativistic. For this type of coupling, axion
emission has no effect in RG stars.
\bigskip

We have also calculated the axion emission through the electron-nucleus
bremsstrah\-lung process, for the case of relativistic, ultra-degenerate
electrons. Our final result is shown in Eq.~(34). For the red giant case,
the density is $\rho=10^6$ g/cm$^3$, so that we have
\be
   \epsilon_a = g_{ae}^2\ 3.8\times 10^{26} \( {T\over 10^8{\rm K}}\) ^4
                \ {\rm erg/g/s}
.\ee
The difference between the non-relativistic result  and the relativistic one
is about a factor 2 in favor of the relativistic case. At $T=10^8$ K,
our result is almost the same as the one used in \cite{DSS}, where a numerical
code was used to set the constraint on the axion coupling. Hence, our
result translates into the following bound
\be
       g_{ae} < 1.5\times 10^{-13}
.\label{RGB}\ee
Should we use the usual limit for the extra energy loss rate that is
tolerable, $\epsilon_x < 30$~erg/g/s \cite{Raf0}, we would get
a less restrictive bound, $g_{ae} < 3\times 10^{-13}$.

For the axion luminosity in white dwarfs and using the same parameters as
in \cite{Raf0}, we find
\be
     L_a = g_{ae}^2\ 7\times 10^{22} L_\odot {M\over M_\odot}
           \( T\over 10^7 {\rm K}\) ^4
,\ee
a result with the same temperature dependence as in previous works, but
somewhat larger \cite{Raf1,Nak}. Again, the origin of the discrepancy lies
in the relativistic effects.

Requiring that the axion luminosity should not exceed the photon
luminosity, we obtain the following bound
\be
     g_{ae} < 1.8 \times10^{-13}
,\ee
which is sligthly weaker than the RG bound.

In the DFSZ model, Eq.~(\ref{RGB}) gives
\be
     m_a < 5.3\times 10^{-3} {\rm eV}/\cos^2\beta
,\ee
which is as good as the SN1987A constraint, with a much lower uncertainty.
In conclusion, we see that the evolution of red giants and white dwarfs
leads to roughly the same constraint on the axion-electron coupling
strength. These results can be summarized in a very conservative bound
\be
     g_{ae} < 2 \times10^{-13}
.\ee
\bigskip

We conclude on the cooling of G117-B15A. The authors of \cite{IHG} have
used a result \cite{Nak} that apparently underestimates the axion emission
rate by a factor 4. Consequently, the value of $g_{ae}$ needed to account for
the
fast cooling of G117-B15A is weaker by a factor of 2 compared than their
estimate. A value as low as $g_{ae}=1.2\times 10^{-13}$ would reconcile
the models with the data. This may have no consequence either on the red giants
evolution or on SN1987A.
\bigskip
\bigskip

{\noindent\large\bf Acknowledgements}\\
We would like to thank G.~Raffelt for many long and helpful mails about
the Primakoff process.

\newpage
{\large\bf Figure Captions\\[0.5cm]}
\begin{description}
\item[Fig.~1] Feynman diagram for the Primakoff process.
\item[Fig.~2] Axion energy loss rate due to the Primakoff effect for
a wide range of densities typical of red giant stars. The solid curve is
our result (Eq.~(11)). Also plotted are Raffelt's result (dashed curve) and the
pole-cut rate with transverse photons only (dotted curve). The value of the
axion-photon coupling has been fixed to $G=10^{-10}$ GeV$^{-1}$.
\item[Fig.~3] The two cut graphs that are needed to evaluate the
bremsstrahlung contribution to the axion emission rate.
\end{description}
\end{document}